\documentclass[rmp,twocolumn,showpacs]{revtex4}

\usepackage{dcolumn,graphicx,amsmath,amssymb,txfonts}

\begin{document}

\title{Detecting degree symmetries in networks}

\author{Petter Holme}
\affiliation{Department of Computer Science, University of New Mexico,
  Albuquerque, NM 87131, U.S.A.}

\begin{abstract}
  The surrounding of a vertex in a network can be more or less
  symmetric. We derive measures of a specific kind of symmetry of a
  vertex which we call \textit{degree symmetry}---the property that
  many paths going out from a vertex have overlapping degree
  sequences. These measures are evaluated on artificial and real
  networks. Specifically we consider vertices in the human metabolic
  network. We also measure the average degree-symmetry coefficient for
  different classes of real-world network. We find that most studied
  examples are weakly positively degree-symmetric. The exceptions are
  an airport network (having a negative degree-symmetry coefficient)
  and one-mode projections of social affiliation networks that are
  rather strongly degree-symmetric.
\end{abstract}


\maketitle

\section{Introduction}

\begin{figure}
  \resizebox*{0.4\linewidth}{!}{\includegraphics{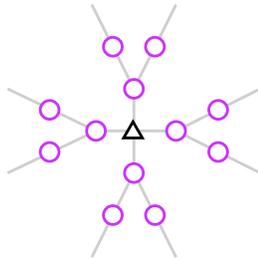}}
  \caption{ Illustrations of degree symmetry. Consider
    paths of length two (i.e.\ $l=2$). All paths out from the
    central (black) vertex have the degree sequence $(3,2)$ meaning
    the central vertex has high degree symmetry.}
  \label{fig:ill}
\end{figure}

With the advent of modern database technology numerous large scale
network data-sets have been made available. This development has
triggered a surge of activity in studies of statistical network
properties~\cite{ba:rev,mejn:rev,doromen:book}. The underlying idea of
these studies is that the network structure (the way the networks
differ from completely random networks) contain some information of the
function, both locally and globally, of the network. Hence a common
theme in these works has been the development of structural measures
to characterize network structure.
In this paper we propose and evaluate a measure of a previously
unstudied network structure---a special case of symmetry we call
\textit{degree symmetry}. In geometry an object
is symmetrical if it is invariant to rotations, reflections, and so
on. In networks, with no given geometrical embedding, these concepts
have to be relaxed. Furthermore, we would like to have a continuous
measure saying not only if a vertex is a local center of symmetry or
not, but also how symmetric the vertex is. The aspect of symmetry we
address is, roughly speaking, that if you look at the object
(network in our case) in different ways from a symmetric vertex it
still looks the same. We process of ``looking'' will in our case be
walking along paths (non-self intersecting sequences of
edges). Furthermore, since degree (number of neighbors) is commonly
regarded as the most fundamental quantity relating a vertex to its
function, we say two vertices ``look the same'' if they
have the same degree. We will thus derive our measure by performing
walks along all paths from a vertex and compare the sequence of
degrees of the vertices along these paths. The situation we have in
mind is depicted in Fig.~\ref{fig:ill}---all paths from the central
vertex have degree sequences starting with $(3,2,\cdots)$, thus the
central vertex is highly degree symmetric.

The rest of the paper is organized as follows: First we give a
detailed derivation of the degree-symmetry coefficient (in two
different versions, appropriate for different needs). Then we evaluate
these on example networks and a biochemical network. Finally we
discuss the average degree symmetry of different classes of real-world
networks.

\section{Derivation of the measure}

We will consider the network represented by a graph $G=(V,E)$ of $N$
vertices, $V$, and $M$ edges, $E$. For a vertex $i$ to have high degree
symmetry it has, as mentioned, to have many paths with the same
sequence of degrees. We will use a cut-off $l$ for the pathlength and
consider only paths of that length. The reason
for this cutoff is threefold: First, in all (with possibly some
curious exception) network processes, a vertex is more affected by its closest
surroundings then vertices further away. Thus one would like to have a
lower weight on the contribution from distant vertices. Second, the
number of vertices $n$ steps away grows fast
with the distance from $i$. For finite networks this means that the
paths soon reach the periphery of the network where unwanted
finite-size effects set in. Third, for computation speed, one benefit
from a cutoff.

\begin{figure}
  \resizebox*{0.8\linewidth}{!}{\includegraphics{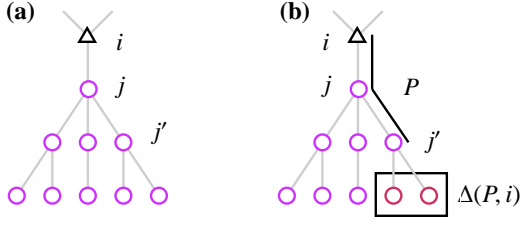}}
  \caption{ Illustrations of of concepts in the derivation of the
    degree symmetry coefficient. (a) illustrates the branching
    number. Consider paths of length three out from $i$. The branching
    number of the path $(i,j)$ is five (there are five paths from $i$
    of length three that goes through $j$). The branching number at
    $j'$ is two. (b) shows the set $\Delta(P,i)$, where $P$ is the
    path $(i,j,j')$.
}
  \label{fig:def}
\end{figure}

Assume there are $p$ paths of length $l$ from a vertex $i$. We then
denote the degree sequences of these paths
\begin{eqnarray}
  Q_l(i)&=&\Big\{[k(v^1_{1,i,l}),\cdots,k(v^l_{1,i,l})],\nonumber\\
&&\vdots\\
 && ,[k(v^1_{p,i,l}),\cdots,k(v^l_{p,i,l})]\Big\},\nonumber
\end{eqnarray}
where $k(v)$ denotes the degree
of a vertex $v$ and $v^j_{m,i,l}$ is the $j$'th vertex of along the
$m$'th path of length $l$ leading out from $i$. Then if there are
unexpectedly many vertices at the same ($j$-) index in the sequence
with the same degree, the vertex $i$ is a local center of degree
symmetry. A rough symmetry measure would thus be to count the fraction
of index-pairs with the same degree, i.e.\
\begin{equation}\label{eq:para}
  \frac{\tilde{s}_l(i)}{\Lambda}=\sum_{0\leq n<n'\leq p}\sum_{j=1}^l
    \delta\big(k(v^j_{n,i,l}), k(v^j_{n',i,l})\big),
\end{equation}
where
\begin{equation}\label{eq:lambda}
\Lambda=(l-1)\:\dbinom{p}{2}\mbox{~~and~~} \delta(x,y)=\left\{
\begin{array}{cl} 1 & \mbox{if $x=y$}\\ 0 &\mbox{if $x\neq
    y$}\end{array}
\right. .
\end{equation}
This measure is very crude and lack many desired statistical
features. For example, all paths that go
via a particular neighbor of $i$ will give a contribution to the
sum. In practice this means that vertices with a high degree
vertex rather far from itself (but closer that $l$) will trivially
have a high $\tilde{s}_l(i)/\Lambda$. A first step would thus be to omit the
contribution of vertices occurring in many sequences of $Q_l(i)$ at a
specific index. I.e., for all $l'\in (0,l)$ one wants to exclude the
terms
\begin{equation}\label{eq:para2}
  \sum_{n,n'} \delta\big(k(v^1_{n,i,l}),
  k(v^1_{n',i,l})\big),
\end{equation}
where $n$ and $n'$ are indices of paths that are identical the
first $l'$ steps, from Eq.~(\ref{eq:para}). Let $S_l(i)$ denote the
number of such terms.

To calculate $S_l(i)$ consider a path $P=(i,\cdots,j)$ of length $l'<l$. Let
$b_l(P,i)$ be the number of paths from $i$ of length $l$ that start
with the path $P$. We call $b_l(P,i)$ the \textit{branching number} of
$P$, see Fig.~\ref{fig:def}(a). All pairs of paths starting with $P$
will contribute to $\tilde{s}_l(i)$ a distance $l'$ from $i$ (since
they all pass
through $j$). Let $\Delta(P,i)$ be the set of neighbors to
$j$ that is not on the path $P$ from $i$ to $j$, see Fig.~\ref{fig:def}(b). (The number of
elements in $\Delta(P,i)$ is thus $k_j-1$.) 
This situation gives a contribution
\begin{equation}\label{eq:cont}
  S_l(P,i) = \dbinom{b_l(P,i)}{2}+ \sum_{j\in\Delta(P,i)}S_l((P,j),i)
\end{equation}
from vertices of indices in the interval $[l',l]$ of $Q_l(i)$ to
$\tilde{s}_l(i)$, where $(P,j')$ denotes the path $(i,\cdots,j,j')$.

To further improve the
measure one would like to, assuming some null-model, subtract the
expected random contribution to $\tilde{s}_l(i)/\Lambda$. If this can be
achieved one would have a symmetry coefficient $s_l(i)$ that is zero when
the symmetry is what can be expected from the null-model, larger if $i$
is a center of unexpectedly high symmetry, and less than zero if $i$ is
degree anti-symmetric. A final symmetry coefficient could thus
be written
\begin{equation}\label{eq:proto}
s_l(i)=\frac{\tilde{s}_l(i)- S_l(i)}{\Lambda- S_l(i)}-\nu , \mbox{~~
  provided $\Lambda > S_l(i)$}
\end{equation}
where $\nu$ is the expected value of $(\tilde{s}_l(i)- S_l(i)) / (\Lambda-
S_l(i))$ in a null-model. $\Lambda= S_l(i)$ can only happen if there is
one or no path of length $l$. In both these cases the degree-symmetry
concept makes no sense so, if $\Lambda =
S_l(i)\in\{0,1\}$, we set $s_l(i)=0$.
The null-model we assume is random
constrained on the degree distribution of the network. I.e., given the
fraction $p_k$ of $k$-degree vertices the network is as random as
possible. As it turns out $\nu$ is tricky to calculate
analytically. There are two ways to proceed---either one calculates
an approximative $\nu$ or one obtains $\nu$ via averaging
$(\tilde{s}_l(i)- S_l(i)) / (\Lambda- S_l(i))$ over realizations of the
null-model. Except being more accurate, the latter approach has the
advantage of giving an error estimate of $s_l(i)$---one can by
specifying a
p-value define significantly symmetric, or anti-symmetric,
vertices. We will use both approaches: The approximative method for
analyzing example networks and the numerical method for
analyzing real-world data.

We obtain an approximative value of $\nu$,  $\nu^\mathrm{app.}$, by
assuming $\nu$ is
approximately equal to the probability that a pair of vertices,
reached by walking along paths, is the same.
Note that, since there are $k$ ways into a
degree-$k$ vertex, when following a path the probability to reach a
degree-$k$ vertex is
\begin{equation}\label{eq:kpk}
  \frac{kp_k}{\sum_{k'} k'p_{k'}} = \frac{kp_k}{\langle k\rangle}.
\end{equation}
Thus the probability $\nu^\mathrm{app.}$ that two vertices of
the same degree is reached by following different paths is 
\begin{equation}\label{eq:nu}
  \nu^\mathrm{app.}=\sum_kp_k\left(\frac{kp_k}{\langle k\rangle}\right)^2 =
  \frac{1}{\langle k\rangle^2} \sum_kk^2p_k^3.
\end{equation}
One reason this approach is not exact is that the number of terms in
the expression for $\tilde{s}_l(i)$ increases with the degree of the
$j$ in $\Delta(P,i)$ of Eq.~(\ref{eq:cont}). There are other
higher-order effects to related to other correlations between the path
structure and the degree of the vertices.

To summarize we have two measures of local vertex symmetry, one
approximative:
\begin{equation}\label{eq:app}
  s^\mathrm{app.}_l(i)=\frac{\tilde{s}_l(i)- S_l(i)}{\Lambda-
    S_l(i)}-\frac{1}{\langle k\rangle^2} \sum_kk^2p_k^3 ,
\end{equation}
and one based one Monte Carlo sampling
\begin{equation}\label{eq:mc}
  s^\mathrm{MC}_l(i)=\frac{\tilde{s}_l(i)- S_l(i)}{\Lambda-
    S_l(i)}-\left\langle \frac{\tilde{s}_l(i)- S_l(i)}{\Lambda-
    S_l(i)}\right\rangle .
\end{equation}
The sampling is conveniently done by random rewiring the edges of the
original network~\cite{roberts:mcmc}.

\section{Algorithm}\label{sect:algo}

The heart of algorithm, as suggested in the previous section, is a
depth-first search with depth $l$. When the returning along the traced
out paths the branching number can be calculated recursively through
\begin{equation}
  b_l(P,i) = \left\{\begin{array}{ll} 1 & \mbox{if $P$ has length $l$}\\
   \sum_{j'\in\Delta((P,j'),i)}b_l((P,j'),i) &  
  \mbox{otherwise}\end{array}\right. . \label{eq:bn}
\end{equation}
$S_l(P_i)$ can be calculated simultaneously using
Eq.~(\ref{eq:cont}). A slight complication is that the same vertex may
appear in different branches of the depth first search while calculating
$b$ and $\tilde{s}$. For small cut-off values this is easy to handle:
For $l=2$ it does not affect the calculation at all. For $l=3$ one
would only have
to keep different depths (of Eqs.~(\ref{eq:cont}) and (\ref{eq:bn}))
separate. For the calculation of $\tilde{s}_l(i)$ the terms
of $Q_l(i)$ has to be stored. Since the number of paths $p$ grows
fast with $l$, this can be quite a constraint for a large
$l$. Luckily it suffices to store a histogram $h(l',k)$ counting the
number of vertices of degree $k$ at position $l'$ of the paths
$Q_l(i)$. $p$ (and thus $\Lambda$) can be calculated as the number of
time the depth $l$ of the depth first search is reached. The running
time of the algorithm is $O(p)$. A mean field
approximation for networks with few triangles gives
$O(p) \approx O(\langle k\rangle^l)$.

\section{Extensions and considerations}

The method outlined above can quite straightforwardly be extended to
network with directed edges, distinct types of edges or (integer) edge
weights.

Imagine a network with $z$ different edge sets $E_1,\cdots,E_z$. Such
networks frequently occur in cellular biochemistry---e.g.\ protein
interaction networks where different types of protein interaction can
be recorded~\cite{hh:pfp}, or gene regulation networks where
the edges can be activating or inhibitory. One sensible way to extend
the above procedure is to use the union of the edges as your
graph but to say two pairs of vertices in $Q_l(i)$ are identical if
their degrees with respect to all of the networks are the
same. To formalize this $Q_l(i)$ would be generalized to
\begin{eqnarray}
  Q_l(i)&=&\Big\{[\mathbf{k}(v^1_{1,i,l}),\cdots,
  \mathbf{k}(v^l_{1,i,l})],\nonumber\\
&&\vdots\\
 && ,[\mathbf{k}(v^1_{p,i,l}),\cdots,
 \mathbf{k}(v^l_{p,i,l})]\Big\},\nonumber
\end{eqnarray}
where $\mathbf{k}(v)$ is a vector with $v$'s degrees with respect to
the different edge-types.
and the $\delta$-function of Eq.~(\ref{eq:para2}) would be one if the
arguments are equal at all their indices, and zero otherwise. The
$\nu^\mathrm{app.}$ has to be redefined too:
\begin{equation}
  \nu^\mathrm{app.} =
  \frac{1}{\langle k\rangle^2} \sum_{k',k''}k'p_{k'}\,k''p_{k''}
  \prod_{i=1}^z\sum_{j=1}^zp_i(k_j|k')p_i(k_j|k''),
\end{equation}
where $p_i(k|k')$ is the conditional probability that a vertex
has degree $k$ with respect to edge set $E_i$ given that its degree
in the union network is $k'$. The case of a directed network can be
treated similarly---one consider paths following edges in both
directions but a vertex pair gives a contribution to $\tilde{s}$ only
if both the in- and out-degrees are the same.

The approach of Sect.~\ref{sect:algo} can straightforwardly be applied
to networks where
multiple edges are allowed. Since multiple edges can be used to model
weighted graphs~\cite{mejn:wei} the generalization to weighted graphs
(at least where edge-weights represent the probability of following an
edge) is simple. The other aspect of multigraphs, self-edges, is
trivially dealt with---by the requirement that a paths should not
intersect themselves a self-edge will never be followed and can thus
be omitted already when the graph is constructed.

The overlap required for a vertex pair to be considered equal in the
calculation of the symmetry coefficient is rather strict. Sometimes one
would like to treat two paths as similar even if their degrees differs
slightly. Particularly, this applies to broad degree
distributions. The functional difference between degree-2 and degree-3
vertices may be significant; but whether a vertex has degree 1002 or
1003 probably does not matter. To achieve such a relaxation one can
construct a integer sequence $K_1<K_2<\cdots$ and let
\begin{equation}\label{eq:ks}
  \delta(k,k') =\left\{\begin{array}{cl} 1 & \mbox{if $K_i\leq
  k,k'<K_{i+1}$ for some $i$}\\ 0 &
  \mbox{otherwise}\end{array}\right. .
\end{equation}
I.e., one construct a series of equivalence classes of vertices.
For a power-law, or similarly broad, degree distributions one can let
$K_{i+1}-K_i$ increase exponentially with $i$. In this case one also
has to modify the definition of $\nu^\mathrm{app.}$
\begin{equation}
  \nu^\mathrm{app.}=\frac{1}{\langle k\rangle^2}\sum_i \left(\sum_{K_i\leq
  k<K_{i+1}} p_k\right) \left(\sum_{K_i\leq k<K_{i+1}} k p_k\right)^2 .
\end{equation}

\begin{figure}
  \resizebox*{\linewidth}{!}{\includegraphics{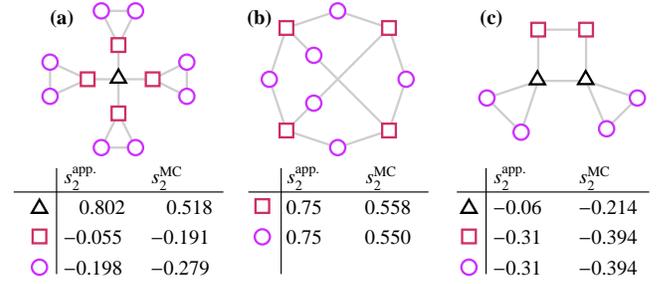}}
  \caption{ Degree symmetries of small example networks. (a) is
    consistent with the example Fig.~\ref{fig:ill}(a). (b) is an
    example of a graph with only positive degree symmetries. (c) shows
    a graph with only negative degree symmetries. The cut-off length
    $l=2$ is used.}
  \label{fig:ex}
\end{figure}

\section{Degree symmetries of example networks}

In this section we evaluate the measure for example networks and
real-world networks. We will use the smallest non-trivial cut-off
$l=2$ throughout this section. Most conclusions hold for $l=3$ or
$4$.

\subsection{Small test graphs}

To get a feeling for the $s_l$ measure we start by considering a few
small test networks, see Fig.~\ref{fig:ex}. In Fig.~\ref{fig:ex}(a) we
display a network with the same degree symmetry, with respect to the
central vertex (triangle), as Fig.~\ref{fig:ill}. As expected the
central vertex has a strong degree symmetry coefficient. To carry
through the calculation of Eq.~(\ref{eq:app}) once we obtain the
degree distribution $p_2=8/13$, $p_3=4/13$ and
$p_4=1/13$ giving $\nu^\mathrm{app.}=165/832\approx0.198$. All length-2 paths
out from the central vertex have the degree sequence $(3,2)$ so
$\tilde{s}_2(\triangle) = 4$, $S_2(\triangle) = 4$ and
$\Lambda= 28$ giving $s_2^\mathrm{app.}(\triangle)=667/832\approx0.802$. The
degree-3 vertices (squares) have two degree sequences of their outgoing
paths $(4,3)$ and $(2,2)$, whereas paths from degree-2 vertices
(triangles) have degree sequences $(3,4)$ and $(2,3)$. This difference
is larger than expected from the null model (random networks with
eight degree-2 vertices, four degree-3 vertices and one degree-4
vertex), thus the negative $s_2$ values for these vertices.

In Fig.~\ref{fig:ex}(b) we show a graph where all vertices have
positive degree-symmetry coefficient. Paths from degree-2 vertices
have only the degree sequence $(3,2)$ and paths from degree-3 vertices
have only the degree sequence $(2,3)$. Thus, for every vertex, the
view of degrees along the path out to the rest of the network is the
same no matter which direction one looks in from that vertex. A
radically different view is seen in Fig.~\ref{fig:ex}(c). In this case
the vertices have three distinct positions in the network. The vertices marked
with squares have degree two and four outgoing paths of degree
sequences $(2,4)$, $(4,4)$, $(4,2)$ and $(4,2)$. The circles, despite
their different network position (as being part of triangles), have
the same set of degree sequences for their paths of length two. The
degree-3 vertices have six length-2 paths: three having the degree
sequence $(2,2)$, three having degree sequence $(4,2)$. It is easy to
convince oneself that this close to as dissimilar a network with four
degree-2 and two degree-4 vertices can be. Consequently all vertices
have negative degree-symmetry indices. It is worth pointing out that
the Fig.~\ref{fig:ex}(c) possesses other symmetries than
degree-symmetry. The layout has, for example, reflexive symmetry along
a vertical axis. We emphasize that such symmetries would need to be
captured by other measures.

\subsection{Regular networks}

If all vertices have the same degree a network is called
\textit{regular}~\cite{janson}. Then by definition all paths are
known to fully overlap. This trivial overlap should be canceled in
our symmetry measure so $s_l(i)=0$ for all $l$ and $i$. Since
$S_l(i)$ is the number of terms in $\tilde{s}_l(i)$ and all
these terms are one we have $S_l(i)=\tilde{s}_l(i)=\Lambda$.
Furthermore, $\nu^\mathrm{app.}=1$ which gives $s_l(i)$ for all
vertices and cut-off lengths.

\subsection{Random graphs}

\begin{figure}
  \resizebox*{0.7\linewidth}{!}{\includegraphics{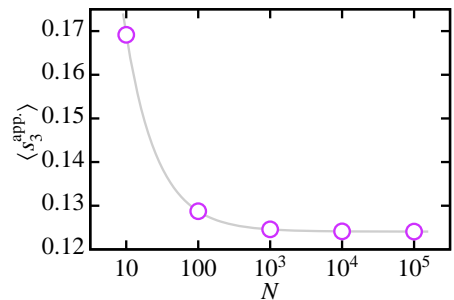}}
  \caption{ The average approximative symmetry coefficient for $l=3$
    and random graphs with $M=2N$. The line is a fit to a power-law
    decay form ($0.124+0.435N^{-1.02}$, to be exact).}
  \label{fig:er}
\end{figure}

Next we evaluate the average approximative symmetry coefficient
$\langle s^\mathrm{app.}\rangle$ for random
graphs~\cite{janson}---graphs obtained by successively adding $M$ edges
between $N$ vertices with the restriction that no multiple edge, or
self-edge, may occur. Such networks have no correlations at all and
can serve as a reference point for neutrality~\cite{mejn:rev}. Ideally
we would like such networks to, on average, have a degree-symmetry
coefficient of zero. As seen in Fig.~\ref{fig:er} $\langle
s^\mathrm{app.}_l\rangle$ converge to a small but positive value.
The decay is roughly inversely proportional to $N$---the same scaling as
the fraction of triangles in the network---which suggests that the
presence of triangles, and perhaps other short-cycles, is an important
source of finite size effects of $s^\mathrm{app.}_l$.
We conclude that the Monte Carlo sampling measure $s^\mathrm{MC}_l$
(or a more elaborate measure) is 
needed if one wants to compare different networks. If, on the other
hand, one aims to compare different vertices of the same network the
faster $s^\mathrm{app.}_l(i)$ calculation is sufficient. This is not
an uncommon situation in the design of network measures. Another
example of this where neutrality is non-zero in the large-$N$ limit is
\textit{modularity}, measuring how good a subgraphs that
are densely connected within but not between each
other~\cite{gui:mod}.

\section{Degree symmetries of real networks}

In this section we apply our measures to real-world networks. First we
take a look at the symmetry coefficients of specific vertices in the
metabolic network of humans, then we look at the average symmetry
coefficients of various classes of networks.

\subsection{Human metabolic networks}\label{sect:meta}

\begin{figure}
  \resizebox*{0.9 \linewidth}{!}{\includegraphics{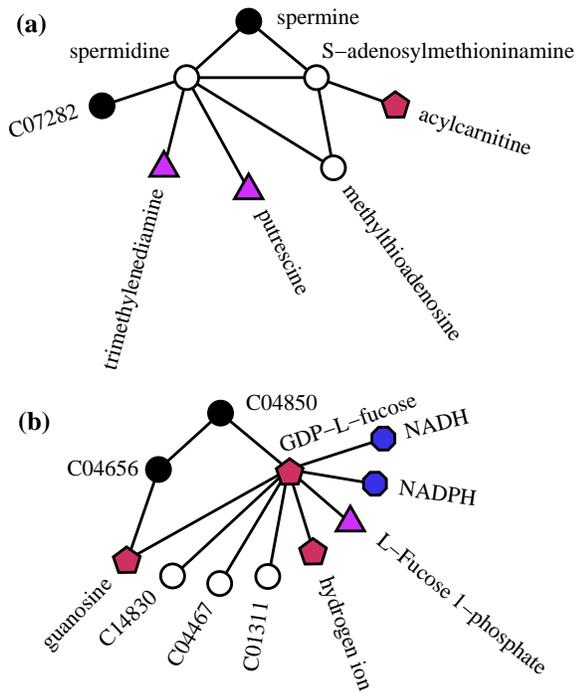}}
  \caption{ The 2-neighborhood of spermine---a vertex with high
    degree-symmetry---(a), and C04850---a vertex with low degree
    symmetry---(b), in the human metabolic network. The symbols
    indicate the equivalence classes
    defined by exponentially growing intervals. Filled circles have
    degree two, unfilled circles have degree four or five, a vertex
    symbolized by  an
    $n$-gon have degree in the interval $[2^n,2^{n+1})$. 
    In case the chemical names are overly long the
    KEGG codes are given (``C'' and five digits):  C07282 represents
    eIF5A-precursor-deoxyhypusine, C04850 represents
    1,3-$\beta$-D-galactosyl-($\alpha$-1,4-L-fucosyl)-N-acetyl-D-glucosaminyl-R,
    C04556 represents 4-amino-2-methyl-5-phosphomethylpyrimidine,
    C04467 represents $\alpha$-L-fucosyl-1,2-$\beta$-D-galactosyl-R
    and C01311 represents
    1,4-$\beta$-D-galactosyl-($\alpha$-1,3-L-fucosyl)-N-acetyl-D-glu\-cos\-aminyl-R. }
  \label{fig:hsa}
\end{figure}

An important use of statistical graph theory is to characterize
chemical reaction networks. Of many possible network
representations~\cite{zhao:meta} we let vertices be chemical
substances, and for all reactions of an organism we link
substrates with products. For example, the hypothetical reaction
$\mathrm{A}+ \mathrm{B} \longleftrightarrow \mathrm{C}+\mathrm{D}$ would
contribute with the edges $(\mathrm{A},\mathrm{C})$,
$(\mathrm{A},\mathrm{D})$ and  $(\mathrm{B},\mathrm{C})$,
$(\mathrm{B},\mathrm{D})$ to the metabolic network. The data is
derived from the KEGG database
(\url{http://www.genome.jp/}), and described in detail in
Ref.~\cite{our:bio}. Since the degree distributions of metabolic
networks are highly skewed~\cite{jeong:meta}
we use a exponentially increasing set of intervals as equivalence
classes (as discussed in the connect of Eq.~(\ref{eq:ks})): $K_n=2^n$.

It has been argued that degree is strongly related to the function of
the chemical substance~\cite{jeong:meta,gui:meta}. This means that the degree
symmetry potentially can give additional information about the function of the vertices. For the
human metabolic network, and $l=2$, roughly half of the vertices have
a p-value of less than 5\% (i.e., in the null-model sampling of the
calculation of $s_2^\mathrm{MC}$, less than 5\% or more than 95\% of
the values of
\begin{equation}
\frac{\tilde{s}_l(i)- S_l(i)}{\Lambda -  S_l(i)}
\end{equation}
are smaller than the value of the real network). In Fig.~\ref{fig:hsa}(a)
we show the 2-neighborhood of one vertex with significantly higher
$s_2^\mathrm{MC}$ than expected; Fig.~\ref{fig:hsa}(b)
depict the 2-neighborhood of a vertex with significantly higher
$s_2^\mathrm{MC}$. The reason these particular vertices are used as examples is
that their 2-neighborhoods are of appropriate sizes, neither too big,
nor too small, to be displayed and described. Spermine,
Fig.~\ref{fig:hsa}(a), is a substance with high
degree-symmetry---$s_2^\mathrm{MC} = 0.89\pm0.02$. Both its neighbors are
in the same degree-equivalence class of vertices with degree four to
seven. Of vertices two steps away from spermine there is also a
significant overlap with two (out of four) neighbors to the neighbor
spermidine being in the equivalence class defined by degrees in the
interval $[8,16)$; whereas two vertices are in the equivalence class
of degrees in
$[4,8)$. The three paths from spermine via S-adenosylmethioninamine
also contribute to the overlap in the two steps from spermine as two
vertices (methylthioadenosine and spermindine) have degrees in the
same equivalence class. The neighborhood of C04850, seen in
Fig.~\ref{fig:hsa}(b), is visually less balanced and also having a
negative degree-symmetry---$s_2^\mathrm{MC} = -0.11\pm0.01$. We note that
there are some vertex pairs in the second neighborhood whose
degree-classes overlap, but apparently this is not enough to make the
symmetry coefficient non-negative.

\subsection{Average symmetry values}

\begin{table*}
\caption{\label{tab:avg} The network sizes $N$ and $M$ and the average
  numerical degree-symmetry coefficient $s_2^\mathrm{MC}$ of
  real-world networks. In the interstate
  network the vertices are American interstate highway junctions and
  two junctions are connected if there is a road with no junction in
  between. In the street networks the vertices are Swedish city-street
  segments connected if they share a junction.
  In the airport network (obtained from
  http://vlado.fmf.uni-lj.si/pub/networks/pajek/data/gphs.htm) the
  vertices are American airports and edges represent a regular, non-stop
  route. In the citation
  networks the vertices are papers and two papers are connected if
  they one cites the other. The ``scientometrics'' network consists of
  papers from the journal \textit{Scientometrics}. The ``small-world''
  network are all papers citing Ref.~\cite{milg:1} or having the
  phrase ``small world'' in the title. (The citation networks were
  obtained from
  http://vlado.fmf.uni-lj.si/pub/networks/data/cite/. These networks
  are the result of searches in the WebofScience used with
  the permission of ISI Philadelphia.) The board of directors and
  Ajou student networks are derived from one-mode projections of
  affiliation networks (where edges goes from persons to corporate
  boards and university classes respectively). The Ajou student
  network is averaged over graphs of 16 semesters. One edge
  represent two students taking at least three classes together that
  semester. The high school networks are gathered from
  questionnaires---an edge means that two persons have listed each
  other as acquaintances. It is averaged over 84 individual
  schools. In the electronic communication networks one edge
  represent that at least one of the vertices has contacted the other
  over some electronic medium. The food webs are networks of
  water-living species and an edge means that one species prey on the
  other. For the protein networks an edge means that two proteins
  interact (the two graphs correspond to two different
  types of experiments determining the interaction edges). The
  metabolic networks consist of chemical substances and edges are
  constructed as described in Sect.~\ref{sect:meta}. Values for
  animal metabolism is averaged over six networks, fungi metabolism
  is averaged over two, and bacteria metabolism is averaged over 96
  networks.
}
\begin{ruledtabular}
 \begin{tabular}{lrc|ccc}
  \multicolumn{2}{c}{network} & Ref. & $N$ & $M$ &
  $s_2^\mathrm{MC}$\\\hline
  geographical networks & interstate highways & & 935 & 1315 &
  $0.016\pm 0.003$ \\
    & streets, Stockholm & \cite{rosv:city} & 3325 & 5100 & $0.014\pm
    0.003$ \\
    & streets, Malm\"{o} & \cite{rosv:city} & 1868 & 3026 & $0.020\pm
    0.003$ \\
    & streets, G\"{o}teborg & \cite{rosv:city} & 1258 & 1516 &
  $0.026\pm 0.003$\\
   & airport & & 332 & 2126 & $-0.0573\pm 0.0002$ \\\hline
    citation networks & scientometrics &  & 2728 & 10398 &
    $0.015\pm 0.020$\\
    & small-world &  & 233 & 994 &
    $0.007\pm 0.002$\\\hline
    one-mode projections of & board of directors & \cite{davis} & 6193 &
    43074 & $0.175 \pm 0.004$ \\ 
    affiliation networks & Ajou University students &
    \cite{our:ajou2}& $7285 \pm 128$
    & $75898\pm 6566$& $0.13\pm 0.01$ \\\hline
    acquaintance networks & high school friendship & \cite{addh} &
    $571\pm 43$ &  $1104\pm60$ &$0.020\pm 0.002$ \\\hline
    electronic communication networks& e-mail & \cite{eckmann:dialog} & 3186 &
    31856 &  $-0.01\pm 0.01$\\
    & Internet community & \cite{pok} & 28295 & 115335
    & $0.01898\pm 0.0001$ \\\hline
    food webs & Little Rock lake & \cite{martinez:rock} & 92 & 960 &
     $0.042\pm 0.001$ \\
    & Ythan estuary & \cite{ythan1} &  134 & 593 &  $0.027 \pm
    0.002$\\\hline
    neural network & \textit{C.\ elegans} & \cite{cenn:brenner} & 280
    & 1973 & $0.0839\pm 0.0001$ \\\hline
    biochemical networks & \textit{S.\ cervisiae} protein &
    \cite{pagel:mips,hh:pfp} & 4580 & 7434
    &  $0.0205 \pm 0.0001$\\
    & \textit{S.\ cervisiae} genetic & \cite{pagel:mips,hh:pfp} & 4580
    & 5129 & $0.0996\pm 0.0001$ \\
    & animal metabolism & \cite{our:bio} & $1621\pm 123$& $4662\pm
    473$ & $0.02\pm 0.01$ \\
    & plant metabolism, \textit{A. thaliana} & \cite{our:bio} & 1561 &
    4302 & $0.0133\pm 0.0003$ \\
    & fungi metabolism & \cite{our:bio} & $1281\pm 97$& $3654 \pm 289$&
    $0.03 \pm 0.02$ \\
    &  bacteria metabolism & \cite{our:bio} & $1070 \pm 35$ & $2776\pm
    109$ & $0.018\pm 0.002$
    \\
  \end{tabular}
\end{ruledtabular}
\end{table*}

So far we have discussed degree symmetries of vertices. In this
section we average $s_l$ over $V$ to obtain
a graph-wide measure for degree symmetry. In Table~\ref{tab:avg} we
display values of
$s_2^\mathrm{MC}$ for a number of different network types. Some of
these have highly skewed degree distributions. For these,
the exponentially increasing degree equivalence classes of
Sect.~\ref{sect:meta} are appropriate. Since we intend to compare all
networks we use the same equivalence classes for all networks. The
first observation is that almost all networks have a positive average
symmetry coefficient. The only clear exception is the airport
network. This means that if you start a two-leg airplane
trip at a particular airport, choosing between two random itineraries
(without caring about the frequency of flights), then the probability
of the airports along these itineraries being different in number of
connections is smaller than in a random network. The strongest
degree-symmetries are found in one-mode projections of social
affiliation networks. Note that the other social networks, derived
from questionnaires and electronic communication does not have such
strong symmetry coefficients.
In one-mode projections high-degree vertices are
known to have strong tendency to attach to other
high-degree vertices, and
low-degree vertices to attach to other low-degree-vertices---so called
assortative mixing~\cite{mejn:assmix}. If this property is strong
there will be regions of vertices with high degree and other regions
with low-degree vertices. The paths within these regions would also
have similar degree sequences. Thus high assortative mixing can be related
to high degree symmetry, the first causing the second or vice
versa. They are, of course, not equivalent---e.g., the example network
with all vertices having positive symmetry coefficients
(Fig.~\ref{fig:ex}(b)) is maximally disassortatively mixed (in the
sense of Ref.~\cite{mejn:assmix}). Where the weak symmetry coefficients
of other networks come from is outside the scope of this
investigation. One possible explanation would be that functional
units~\cite{alon} might often be degree-symmetric centers.

\section{Summary and conclusions}

We have derived a measure for a specific notion of symmetry in
networks---the property that the paths out from a vertex have
overlapping degree sequences. The measure is designed so that random
networks, conditioned only to have the same set of degrees as the
original network, have the value zero. We propose two versions of the
symmetry coefficient, the first being approximately zero for random
networks, the second requiring a randomization procedure (and thus
longer simulation time) but being more accurately zero for random networks. The measure was
evaluated on example graphs. We show that they are able to
detect vertices in degree-symmetric, and potentially functionally
meaningful positions in the human metabolic network. The average
degree-symmetry of various networks were also investigated. We found almost
all networks having a weakly positive degree coefficient. The
exceptions being the network of American airports and their
interconnections (having a negative degree-symmetry coefficient) and
one-mode projections of social affiliation networks (having rather
strongly positive values).
Our measure is not the first to be based on a the properties of paths
going out from a vertex. For example people have been using path
counts for assessing the functional similarity of pairs of
vertices~\cite{blondel:sim,simrank,our:sim}. In social network studies
such measures are commonly called ``ego-centric''~\cite{wf}.

Symmetry concepts have been successfully utilized in many field of
physics. We believe degree symmetry, and other classes of network
symmetries, will be a fruitful direction of future network
studies. Degree symmetry is in particular, we believe, an important
concept for networks where degree is strongly related to the function
of the vertex. Two open questions from this study is what causes the
rather ubiquitous weakly positive degree symmetries, and what process in the
airline decision making that causes the negative average symmetry
coefficient of the airline network.

\begin{acknowledgements}
  The author acknowledges financial support from the Wenner-Gren
  foundations and help with data acquisition from: Gerald Davis,
  Jean-Pierre Eckman, Michael Gastner, Mikael Huss, Beom Jun Kim,
  Sungmin Park and Martin Rosvall. This research uses data from Add
  Health, a program project designed by J. Richard Udry, Peter
  S. Bearman, and Kathleen Mullan Harris, and funded by a grant
  P01--HD31921 from the National Institute of Child Health and Human
  Development, with cooperative funding from 17 other
  agencies. Special acknowledgment is due Ronald R. Rindfuss and
  Barbara Entwisle for assistance in the original design. Persons
  interested in obtaining data files from Add Health should contact
  Add Health, Carolina Population Center, 123 W. Franklin Street,
  Chapel Hill, NC 27516--2524 (addhealth@unc.edu).
\end{acknowledgements}

\end{document}